\documentclass[12pt,preprintnumbers,amsmath,amssymb,showpacs,aps,prl,onecolumn,superscriptaddress]{revtex4-1}

\usepackage{graphicx,stmaryrd,tipa,amsmath,amssymb,float,array,color,tabularx,threeparttable,booktabs,bm}
\usepackage{graphicx,color}
\usepackage[colorlinks,citecolor=blue]{hyperref}
\newcommand{\etal}{\textit{et al.\/}}

\begin{document}
\title{The mechanism and process of spontaneous boron doping in graphene in the theoretical perspective}
\author{Xiaohui Deng} \email{x.deng@hynu.edu.cn}
\affiliation{Department of Physics and Electronic Information Science, Hengyang Normal University, Hengyang 421008, People's Republic of China}
\author{Jing Zeng}
\affiliation{Department of Physics and Electronic Information Science, Hengyang Normal University, Hengyang 421008, People's Republic of China}
\author{Mingsu Si}
\affiliation{Key Laboratory for Magnetism and Magnetic Materials of the Ministry of Education, Lanzhou University, Lanzhou 730000, People¡¯s Republic of China}
\author{Wei Lu}
\affiliation{University Research Facility in Materials Characterization and Device Fabrication, The Hong Kong Polytechnic University, Hong Kong, People's Republic of China}

\date{\today}
\begin{abstract}
A theoretical model is presented to reveal the mechanism of B doping into graphene in the microwave plasma experiment choosing trimethylboron as the doping source (ACS NANO {\bf 6} (2012) 1970). The results show that the reason for B doping comes from the combinational interaction of B and other groups (C, H, CH, CH$_2$ or CH$_3$) decomposing from trimethylboron and the doping undergoes two crucial steps. The minimal energy path for the first step are determined. The obtained energy barrier of considered cases fall into the range of 0.02-0.43 eV, supporting the fact that the substituting B for C can easily realized even at room temperature. As the second step, after removing irrelevant groups in vertical direction through H saturation, the perfect B doping is realized at last. This work successfully explain the above experimental phenomenon and propose a novel and feasible method aiming at B doping of graphene.

\end{abstract}



\maketitle
\section{Introduction}
Graphene has been under investigation for the theoreticians and experimentalists due to the novel physical, chemical properties \cite{graphene1,graphene2,graphene3,graphene4,graphene5} and the potential applications in many fields \cite{app1,app2,app3,app4,app5} from its experimental discovery in 2004. \cite{discovery} Graphene is a semimetal with zero band gap, which limits the actual applications of graphene as electron devices. In order to widely apply graphene to the field of electronics in the present Si-based semiconductor times, the first step is to open a band gap in the order of 1 eV for graphene. Several methods have been reported used to open the gap of graphene. The chemical doping, external electric field and molecular adsorption are the suitable and effective methods. \cite{formation-enerngy1,formation-enerngy2} Doping may tune the physical and chemical properties of graphene, however, is known to be a relatively difficult task due to the existing strong $sp^2$-bond. From the theoretical calculations, the formation energy of B doping in bigraphene and graphene are determined to be 5.6 and 13.50 eV, respectively,\cite{formation-enerngy1,formation-enerngy2} suggesting the B doping into graphene really is a challenge. The modulation of doping type and concentration in nanoscale may be very important for the application of graphene. It is a well-known fact that B-doped graphene behaves as a $p$-type semiconductor, promising potential applications as a part of $p$-$n$ junction.

Several techniques were reported to incorporate heteroatoms into graphenes through physical and chemical ways, such as arc discharge, \cite{formation-enerngy1} electrothermal reactions \cite{electothermal} and chemical vapor deposition.\cite{CVD1} However, many doping issues, especially control of doping concentration and corresponding transport properties, remain largely unresolved, so it is difficult to quantify the doping effects yet. The plasma treatment usually also can achieve the aim of doping. Tang \etal \cite{tang} reported a method to control the boron (B) doping concentration into graphene {\it via} controllable doping through reaction with the ion atmosphere of trimethylboron decomposed by microwave plasma. They found that the band gap was tunable and B-doped graphenes had the $p$-type transport property. Theoretically there are many papers aiming to the doping of graphene, wherein the considered dopants coveres from the light atoms to metal atoms, such as B, N, O, F, P, Al, Mn, and Cr. \cite{formation-enerngy2,DG1,DG3,DG4,DG5} However, all of them only focused on the physical or chemical properties after the dopant already doping into graphene. The physical insight on why these dopants can be doped into graphene has not been given, making these doping ideas stay in the theoretical level.

In Tang's experiment,\cite{tang} the trimethylboron B(CH$_3$)$_3$ (TMB) was chosen as the doping source, which has been widely used for $p$-type doping for else materials.\cite{TMB1,TMB2,TMB3} The dissociation energy of B-C bond in TMB is approximately 87 kcal/mol (3.77 eV).\cite{TMB1} As a result, under the lash of the low energy electron cyclotron resonance-enhanced microwave plasma, the TMB molecule is easy to decompose {\it via} the sequential loss of methyl group (CH$_3$). The gaseous boron atoms also is formed. With the continued effect from microwave plasma, the methyl group may be further decomposed. The final composition of TMB may consist of gaseous B, C, CH, CH$_2$, CH$_3$, and H group. At the pressure of 30 Torr and the temperature of 300$^{\circ}$C, the B-doped graphene was prepared after reaction with the ionic atmosphere of TMB. Why and how B dopant be doped into graphene?  In this paper, using {\it ab initio} calculations, we propose an available method for B doping in graphene and successfully explain the experimental B doping process. In our model, B doping can be realized by substituting spontaneously C atom in graphene after two crucial steps.

\section{Computational details}
The geometries exploring and total energy calculations are performed by Vienna Ab-initio Simulation Package (VASP) \cite{vasp1,vasp2} within the framework of pseudopotential plane waves method. The generalized gradient approximation (GGA) proposed by Perdew \etal \cite{pbe} is used for the exchange-correlation functional. The projector-augmented-wave (PAW) \cite{paw} is used to describe the interaction between ions and electrons for C, B, and H atoms. The kinetic energy cutoffs of 500 eV for the plane wave expansion is adopted. A 50-atom supercell is used to simulate a $5\times5$ graphene sheet. The periodical boundary conditions (PBC) are employed in the vertical direction of the sheet and adjacent sheet are separated by a vacuum region of 15 {\AA}. The Brillouin zone sampling chooses a $5\times5\times1$ Monkhorst-Pack\cite{MK} grid in geometries relaxations. To obtain accurate total energy and charge density, the denser $k$-points of $11\times11\times1$ and tetrahedron method \cite{tetrahedron} were chosen. We have optimized all geometries studied here by reducing the Hellman-Feynman forces of every atoms down to 0.02 eV/{\AA}.In order to quantify the stability and adsorption ability of the systems, we calculate the adsorption energy defined as $E_a=E_{t}-E_{p}-E_{g}$, where $E_{t}$ is the energy of graphene with adsorbed groups,  $E_{p}$ is the energy of pristine graphene, $E_{g}$ is the energy of adsorbed groups.

\section{Results and discussion}
It's because that the final composition of TMB may include gaseous B, CH, CH$_2$, CH$_3$, C and H groups, we first clarify the adsorption behaviours of these groups solely adsorbed on the graphene sheet. The preferred adsorption site usually is the high-symmetric sites. Therefore, three classic high-symmetric sites in graphene are considered here. They are the top site (above the C atom), the hollow site (above the center of the hexagon of C atom), and the bridge site (above the middle of two adjacent C atom). The structural parameters and adsorption energies are listed in table {\ref{table1}}. From the table, for B adsorption, the bridge site is the most stable one with the adsorption energy of -0.99 eV, indicating a chemical adsorption behaviour and supported by the classical chemical bond length of 1.68 (1.79-0.11) \AA. After B adsorption, the adjacent C atoms bonded to B atom is slightly displaced from graphene plane (0.11 \AA). The second and third preferred adsorption site are top and hollow sites, respectively. However, the difference in relative energy and adsorption energy is small. For C, CH and CH$_2$ groups, as in B case, the bridge site also is  the most stable site, judging from the bond length ($d_{a}$) and the adsorption energy $E_{a}$. Moreover, the priority of adsorption for B, CH, CH$_2$, and CH$_3$ groups all are in the order of bridge, top and hollow sites. However, H and CH$_3$ trend to adsorb in the top site. It is worth mentioning that the C, CH, CH$_2$, CH$_3$, and H groups can "pull" the C atom out of the honeycomb graphene sheet(around 0.5 \AA), which means these five groups will strongly interact with the graphene plane and disturb the strong $sp^2$-hybrid bond in the graphene lattice. Therefore, it is a change to realize the another elemental doping, for example B, into graphene.

\begin{figure}
\begin{center}
\includegraphics[width=14cm]{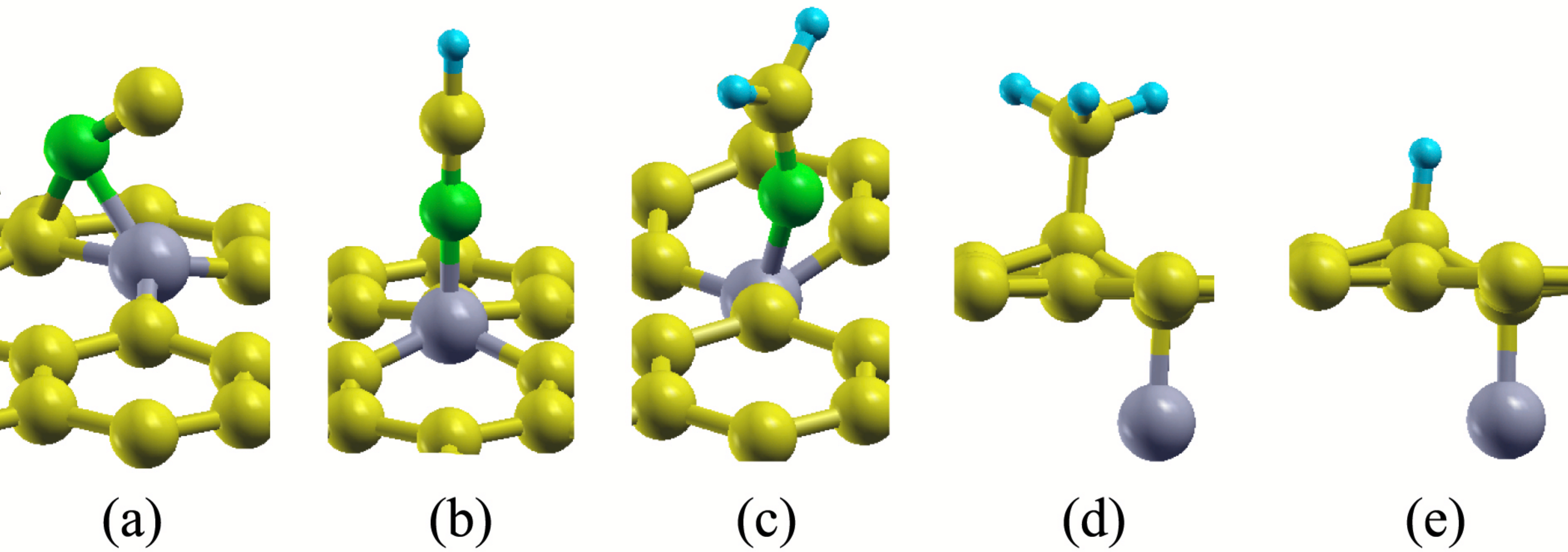}
\caption{(Color online) The relaxed structures when B and C or other derivations from TMB are placed in two surfaces of graphene. (a) the C case, (b) CH, (c) CH$_2$, (d) CH$_3$ and (d) H. The yellow, grey, blue balls express C, B and H atoms, respectively. For clarity, the pulled C atom is picked out as green ball.}
\label{fig1}
\end{center}
\end{figure}

\begin{figure}
\begin{center}
\includegraphics[width=16cm]{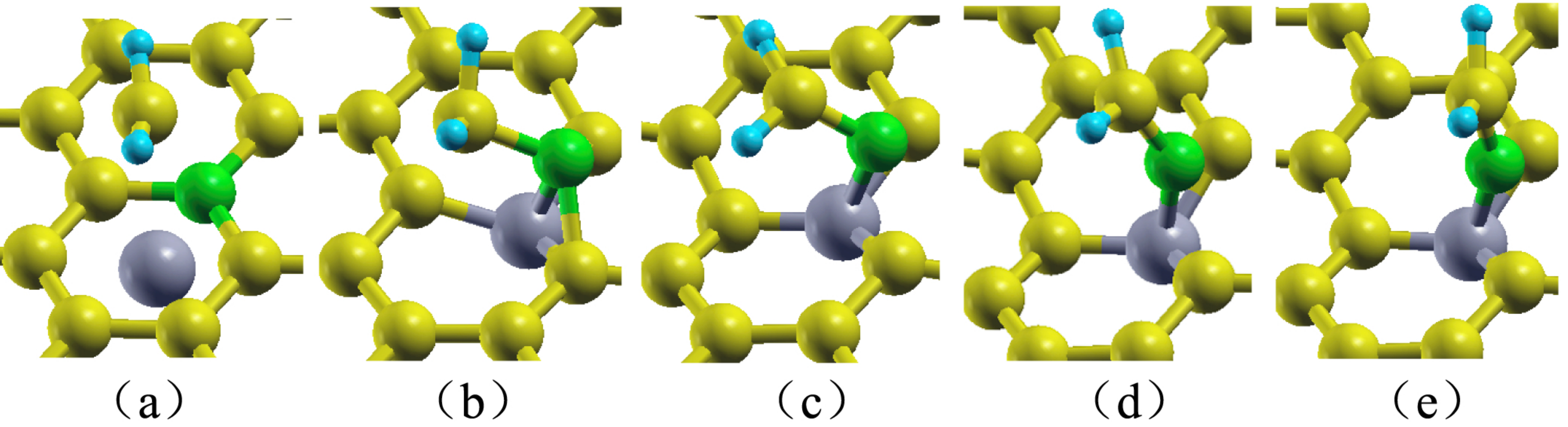}
\caption{(Color online) The process of B doping into graphene with the help of CH$_2$. Where, (a) is the initial structures. B atom and CH$_2$ group are placed at 3.0 {\AA} from the graphene sheet at the beginning. (e) is the final structure with B atom occupying the C site in graphene. The yellow, grey, blue balls express C, B and H atoms, respectively. For clarity, the pulled C atom is picked out as green ball.}
\label{fig2}
\end{center}
\end{figure}

Now we turn our attention to a curious phenomenon. When B and C (or other derivations from TMB) both adhere to their preferred adsorption sites in each side of the graphene sheet, what will happen? In order to simulate the possible process, we construct the initial structures as follows. The B atom is placed in the bridge site of one side of graphene sheet, however, the C (or CH and CH$_2$ group) is placed in the corresponding bridge site of another side of the sheet (top site for H and  CH$_3$ ). The initial distances of B, C, CH, CH$_2$, CH$_3$ and H away from the sheet is 3.0 \AA. The conjugate gradient minimization scheme is used to minimize forces on atoms until all of the force components obeys the self-consistent criterion. The relaxed structures are shown in Fig. \ref{fig1}. One can find that under the pull force from C, CH and CH$_2$, one C atom is pulled out of the graphene plane with B atom substituting its site. However, the effect by using H and CH$_3$ is not obvious. That is to say, the B doping will be spontaneously doped into graphene under the help of C, CH and CH$_2$ to satisfy energy minimization, only depending on the minimization process of force. The H or CH$_3$ group seat on the top of one C atom, the B atom locates on the top of another C atom. It is easy to understand. The CH$_3$ and H groups possess one valence electron. Therefore, they favor to bond with a C atom with valance style. The B atom moves from bride site to top site to reduce its energy. The situation is similar to two H adsorbed graphene.\cite{HonG}  In order to clearly observe the phenomenon of B substituting C, take CH$_2$ for example, we illustrate the process in Fig. \ref{fig2}. One can find that: (1) the CH$_2$ group tends to pull one C atom (green ball) out of the graphene plane and (2) B atom trends to bond with three adjacent C atoms around the green C atom. The strength of C-C bonds are gradually weaken due to the combination of such two reasons. The green C atom is pulled from its original site and is replaced with B atom at last. In final structure, the average C-B distances in plane is found to be 1.52 {\AA}. In vertical direction of the sheet, $d_{B-C}$ is 1.67 {\AA}, indicating the C$_2$H$_2$ group chemically bonds with the sheet. In order to realize the perfect B doping, the vertical C$_2$H$_2$ group must to be removed.

\begin{figure}
\begin{center}
\includegraphics[width=16cm]{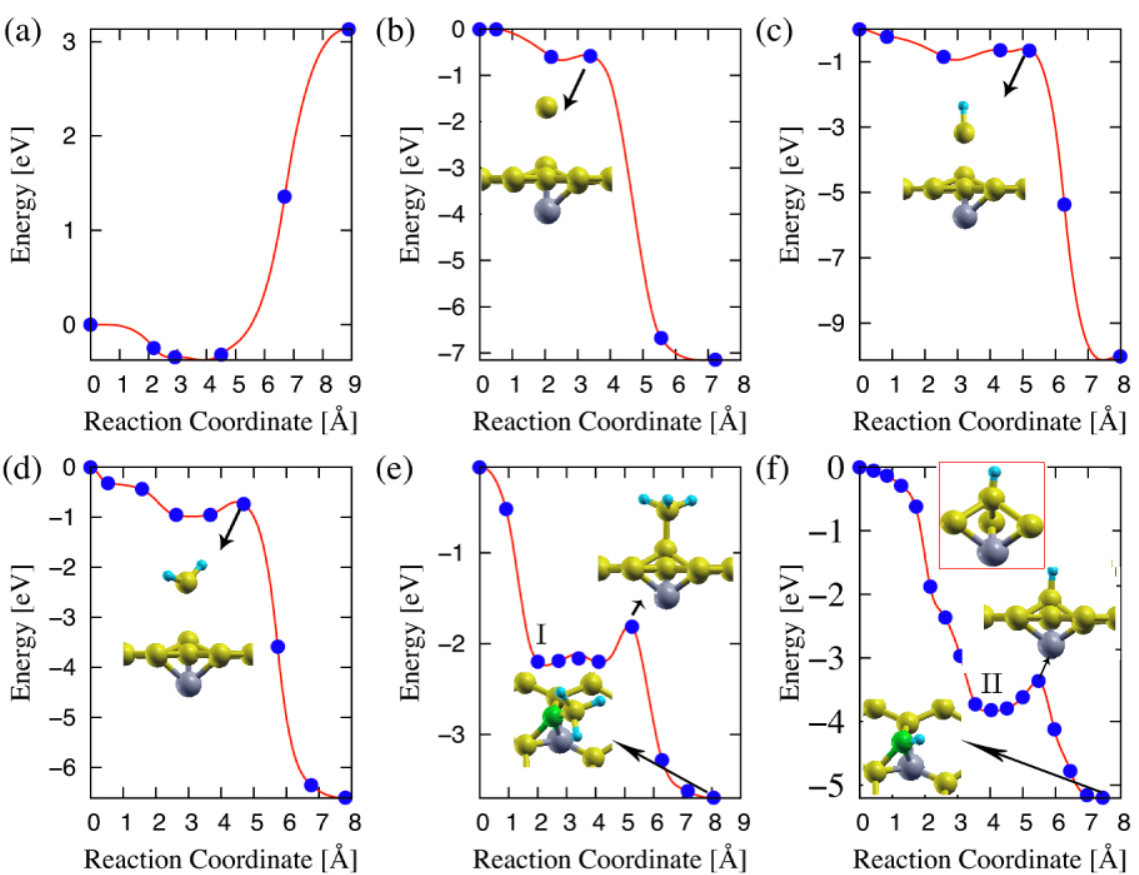}
\caption{(Color online) The minimal energy paths (MEP) of B doping into graphene for (a) a B atom far away from the graphene sheet substituting a C atom withhout the help of derivations from TMB, (b) with the help of C, (c) CH, (d) CH$_2$, (e) CH$_3$, and (f) H case. The energy barrier for six processes are 3.13, 0.02, 0.18, 0.22, 0.40, 0.43 eV, respectively. The structures for I and II are Fig. \ref{fig1} (d) and (e). The reduced unit like ditetrahedral structure is inserted in the red rectangle in (f) after removing irrelevant atoms}
\label{neb}
\end{center}
\end{figure}

In order to explain this spontaneous behaviour, the climb image nudged elastic band (CI-NEB) method \cite{neb1,neb2} is used to search the minimal energy paths (MEP) for considered processes. The B and derivations of TMB initially are placed in their favored sites far away the graphene sheet enough. Between the initial and final configurations, there are 4-16 images on demand for the CI-NEB calculations. Without the combination of derivations from TMB, the B doping into graphene will overcome a barrier of 3.13 eV, as shown as in Fig. \ref{neb} (a), which directs out that the substitutional B doping truly is of relatively difficult. The formation energy $E_f$, defined as in our previous work,\cite{jcp2012} also are adopted to estimate the formation ability of B-doped graphene. The yielded $E_f$ is equal to 2.65 eV, being consistent with the value from CI-NEB calculation. With the help of C, CH and CH$_2$, C substituting with B atom override the barriers of 0.02, 0.18, and 0.22 eV, respectively. These low reactive barrier is insufficient to prevent the B atom beating the C atom. The MEP for the CH$_3$ and H cases show two barriers being 0.40 and 0.43 eV, respectively, which results in that the local minimal energy (denoted by I and II for CH$_3$ and H) are explored to be the structures as Fig. \ref{fig1} (d) and (e) during the structure relaxation process. However, such two barriers are easily to overcome even at room temperature with taking into account temperature effect on ions. It is worth pointing out that the transition states for all five cases ((b)$\rightarrow$(f)) are ditetrahedral structure (insert of Fig. \ref{neb} (f)) despite the  barrier energy being big or small. A conceivable factor may be that the electronegativity is in the order of C, CH, CH$_2$, CH$_3$, and H. The interaction between graphene and the ditetrahedral unit gradually decreases as the electronegativity decreases until the H and CH$_3$ have no enough electron to disturb the bonds in ditetrahedral unit.

\begin{figure}
\begin{center}
\includegraphics[width=16cm]{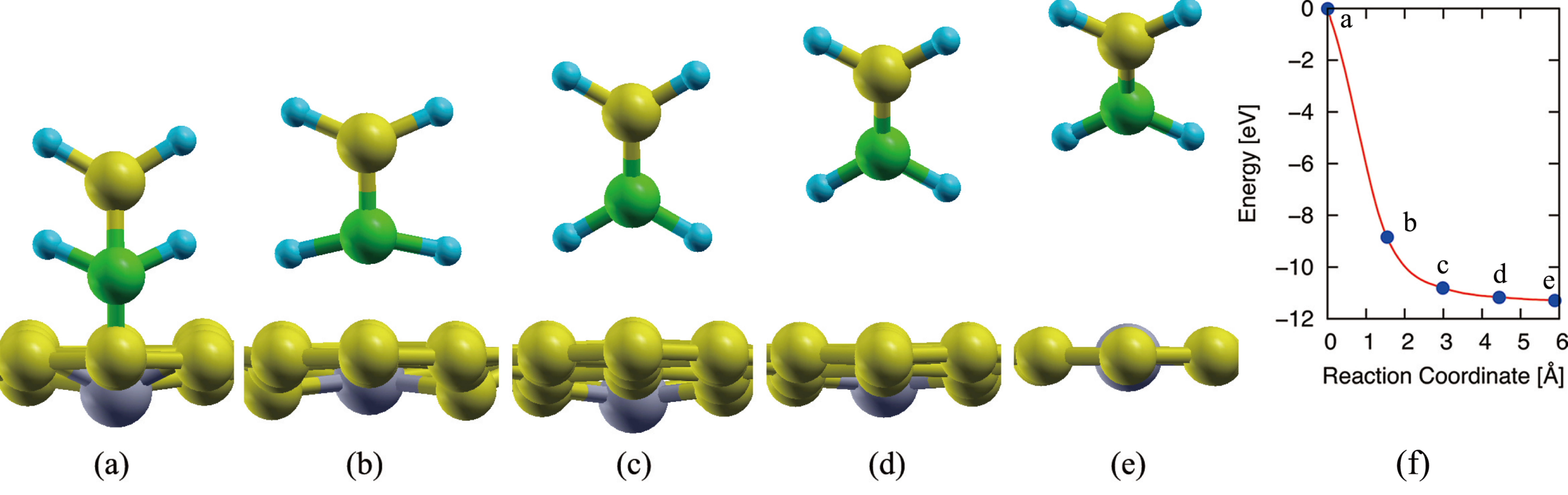}
\caption{(Color online) The removing process of -CH$_2$ group. (a) is the initial structures (the same structure as Fig. \ref{fig2}(e), but in different view direction). Two H atoms are arbitrarily placed beside the green C atom.  The two H atoms can saturate the CH$_2$ group, forming C$_2$H$_4$ molecule ((b-(e))). The yellow, grey and blue balls express C, B and H atoms, respectively. For clarity, the pulled C atom is picked out as green ball.(f) is the MEP for the above process.}
\label{fig4}
\end{center}
\end{figure}

As mentioned above, although B can be doped into graphene, the pulled C atom and C, CH, and CH$_2$ from TMB are chemically bonded to the B-doped graphene sheet in vertical direction (Fig. \ref{fig1} (a)-(c)). The vertical adsorbed groups must to be removed to arrive in the aim of perfect B doping. In derivations of TMB have enough atomic H, which can be seen from the dissociation reaction described as B(CH$_3$)$_3$$\rightarrow$B+3(C+3H), B(CH$_3$)$_3$$\rightarrow$B+3(CH+2H), and B(CH$_3$)$_3$$\rightarrow$B+3(CH$_2$+H). These extra H can be used to remove the vertical adsorbed groups through saturating its chemical bonds. Take CH$_2$ for example, we arbitrarily place two atomic H beside the C$_2$H$_2$ group. The interesting process is found and shown in Fig. \ref{fig4}. Because of the existence of the two H atoms, the C$_2$H$_2$ group tends to bond with the two H atoms. The bond between B and green C is finally broken ((a)$\rightarrow$(b)). The $\pi$ bond of the formed C$_2$H$_4$ molecule and of B-doped graphene repulse each other. The C$_2$H$_4$ molecule detaches gradually from the sheet ((b)$\rightarrow$(e)), leaving the B-doped graphene alone with a distance of 3.95 {\AA}. The result from MEP calculation, as shown in Fig. \ref{fig4} (f), shows that the removing of CH$_2$ group is energetically preferable. With the distance between C$_2$H$_4$ and B-doped graphene sheet increasing, the energy of system decreases steeply. The desorption energy is about -11.28 eV, indicating such dissociation process is easy to happen. B and C are neighbours in the periodic table of elements. They have comparable ionic radius and electronegativity. Therefore, there is no local distortion after B doping into graphene, being in good agreement with the theoretical work. \cite{formation-enerngy2,DG4,DG5} The B-C distance ( 1.48 {\AA}) is slightly smaller than the C-C distance (1.42 {\AA}), which can be attributed to the fact that the electronegativity of B is lower than that of C, leading to a lower binding energy of the B-C bonds compared to that of C-C bonds.\cite{app1,formation-enerngy2} The C$_2$H$_4$ molecule keeps above the B-doped graphene sheet with a physisorption force. The adsorption energy is about 50 meV. For such weak interaction, the thermal effect even can lead to the detach of C$_2$H$_4$ from the sheet.

Our B doping mechanism is based on the cooperation of B and TMB derivations by placing them in opposite sides of the graphene sheet. The first principle calculations indicate the C-B replacement will occur driven by the combinational interaction from the pull force of TMB (or B) and the push force of B (or TMB). Obviously, the model needs B or TMB derivations can reach the side facing to the substrate. Is our model suitable for the case when B and TMB derivations only can reach the same side of graphene sheet?  We have examined the case with placing B and TMB derivations on the same side of graphene. The results show that C-B replacement does not occur. This is because that B and TMB derivations will compete with each other in binding with C atoms, thus form complex structures. Actually, B or TMB groups can reach the side facing to the substrate. In Tang's experiment, graphene samples were placed on the stage of SiO$_2$/Si. The SiO$_2$ layer is prepared by oxidizing the Si substrate. Therefore, the surface of SiO$_2$ is very rough, where, the wrinkle height may approach to several microns, In additional, the SiO$_2$ surface also exist interface of polycrystal. The above concerning factors induce big interspace to enter for B or TMB groups.

Before reaction with graphene samples, the TMB already is decomposed into ion gaseous compound in Tang's experiment. These derivations with high energy play an important role in overcoming high barriers during the C-B replacement process. The fragments at high-energy state have strong oxidizability, which can disturb the $sp^2$ bond in graphene through exchanging its unsaturation electrons with graphene. The stronger electronegativity of fragment is, the smaller barrier will be.The B doping mechanism actually dependents on the collaborative effect of B and TMB derivations. Without any of them, the goal of B doping cannot achieve. Although the fragments start from a high-energy state, it isn't easy to overcome formation energy of for one C vacancy (7.5 eV). However, the formation energy of B-doped graphene is about 2.65 eV (NEB give a barrier of 3.13 eV), comparable to the dissociation energy of B-C bond in TMB (3.77 eV). Just because of this reason, the replacement of C with B occurs at last. The mechanism of B doping in this paper also may be suitable for other 2D carbon materials if one doping process has matching or lower reaction barrier energy comparing with 3.77 eV. 

\section{Conclusion}
In summary, by constructing the same conditions with the microwave plasma experiment (ACS NANO {\bf 6} (2012) 1970), we present a theoretical model to explain the experimental phenomenon of B doping in graphene based on {\it ab-initio} calculations. We find that the detailed process of B doping consisted of two crucial steps. As the first step, with the help of C, CH, or CH$_2$ groups, B atom could substitute the C atom in the graphene plane. However, two other derivations, i.e. CH$_3$ and H, can not help B atom to substitute C atom in graphene seen from the structure relaxation process at zero temperature. The CI-NEB calculations has done to evaluate the creative barrier and give the barrier energy of C-, CH-, CH$_2$-, CH$_3$- and H-induced B doping are 0.02, 0.18, 0.22, 0.40 and 0.43 eV, respectively. It indicate that the aim of B doping will be arrived in with the help of C, CH, and CH$_2$ groups. On the contrary, CH$_3$ and H groups is unpractical due to their relatively big energy barriers. However, taking into account temperature effect on ions, such two barriers are easily to overcome even at room temperature. The reason for different results may be close to the electronegativity of these groups. After removing irrelevant groups, the perfect B doping is realized at last. The proposed method is novel and feasible for B doping of graphene.

\section{Acknowledgments}
We gratefully acknowledge financial support from the Nation Science Foundation of China under grant No. 11304087 and 61401151, the Science and Technology Project of Hengyang
City under grant No. 2013KJ33, the Science Research Fund of Hunan Provincial Education Department of China under grant No. 12A020, and the Construct Program of the Key Discipline in Hunan province of China.

\clearpage
\begin{table}
\caption{Geometric and energetic parameters for the stable adsorption of different groups on the graphene sheet at three classic adsorption sites (the relative energy($E_r$), the distances between adsorbate and the sheet($d_a$), the distances of carbon displaced from the the sheet ($d_c$), and  the adsorption energy $E_{a}$). }
\begin{tabular}{ccccccccccccccc}
\hline \hline
 groups   &        & top site    
       &bridge site       &hollow site      \\
\hline
B    &  $E_{r}$(eV)   & 0.08   & {\bf0.00}  & 0.12    \\
     &  $d_{a}$({\AA})   & 1.76   & 1.79 & 1.64    \\
      &  $d_{c}$({\AA})    & 0.00      & 0.11  & 0.00    \\
      &  $E_{a}$(eV)      & -0.91      & -0.99  & -0.87   \\
C    &  $E_{r}$(eV)   & 0.65   &  {\bf0.00} & 1.52    \\
     &  $d_{a}$({\AA})   & 1.91   & 1.77 &  3.50  \\
      &  $d_{c}$({\AA})    & 0.15      & 0.46  & 0.00    \\
      &  $E_{a}$(eV)      & -0.89      & -1.54  &  -0.02  \\
CH    &  $E_{r}$(eV)   & 1.20   & {\bf0.00}  & 2.20    \\
     &  $d_{a}$({\AA})   & 2.03   & 1.75 & 3.24   \\
      &  $d_{c}$({\AA})    & 0.53      & 0.55  & 0.00    \\
      &  $E_{a}$(eV)      & -1.09      & -2.30  & -0.10   \\
CH$_2$    &  $E_{r}$(eV)   & 1.24   & {\bf0.00}  & 1.87    \\
     &  $d_{a}$({\AA})   & 2.09   & 1.84 & 3.29   \\
      &  $d_{c}$({\AA})    & 0.56      & 0.55  & 0.00    \\
      &  $E_{a}$(eV)      & -0.84      & -2.09  & -0.22   \\
CH$_3$    &  $E_{r}$(eV)   & {\bf0.00}   & 0.20  & 0.20   \\
     &  $d_{a}$({\AA})   & 2.08  & 3.62 & 3.50   \\
      &  $d_{c}$({\AA})    & 0.49      & 0.00  & 0.00    \\
      &  $E_{a}$(eV)      & -0.22      & -0.02 & -0.02   \\
H    &  $E_{r}$(eV)   & {\bf0.00}   & 0.81  & 0.82    \\
     &  $d_{a}$({\AA})   & 1.61   & 3.17 & 3.12   \\
      &  $d_{c}$({\AA})    & 0.48      & 0.00  & 0.00    \\
      &  $E_{a}$(eV)      & -0.84      & -0.03  & -0.02   \\
\hline \hline
\end{tabular}
\label{table1}
\end{table}

\clearpage

\clearpage

\clearpage
\clearpage

\end{document}